\documentclass[journal]{IEEEtran}
\usepackage{etex}
\usepackage{tikz}
\usepackage[utf8]{inputenc}
\usepackage{pgfplots}
\usetikzlibrary{plotmarks}
\usepackage{pdflscape}
\usepackage{microtype}
\pgfplotsset{compat=newest}
\pgfplotsset{plot coordinates/math parser=false}
\newlength\figureheight
\newlength\figurewidth
\usepackage{epsf,epsfig,array}
\usepackage[cmex10]{amsmath}
\usepackage{amssymb}
\usepackage{amsfonts}
\usepackage[font=footnotesize]{caption}
\usepackage{cite}
\usepackage{graphicx}
\usepackage{epstopdf}
\usepackage{subfigure}
\usepackage{epsf,epsfig}
\usepackage[]{latexsym}
\usepackage{algorithm, algorithmic}
 \usepackage{color}
\usepackage{setspace}
\usepackage{flushend}
\usepackage{lipsum}
\usepackage{multicol}
\usepackage{float}
\usepackage{authblk}
\usepackage{geometry}
\newgeometry{left=0.7in,right=0.35in,top=0.6in,bottom=0.45in}

\newtheorem{proposition}{\emph Proposition}

\makeatletter
  \def\vhrulefill#1{\leavevmode\leaders\hrule\@height#1\hfill \kern\z@}
\makeatother

\ifCLASSINFOpdf
\else
\fi
\begin{document}


\title{\LARGE On the Sum of Fisher-Snedecor $\mathcal{F}$ Variates and its Application to Maximal-Ratio Combining}
\author{{Osamah. S. Badarneh, \emph{Member, IEEE}\thanks{O. S. Badarneh is with University of Tabuk, Electrical Engineering Department, Tabuk, Saudi Arabia (e-mail: obadarneh@ut.edu.sa).}, Daniel B. da Costa, \emph{Senior Member, IEEE}\thanks{D. B. da Costa is with the Department of Computer Engineering, Federal University of Cear\'{a}, Sobral, CE, Brazil (e-mail: danielbcosta@ieee.org).}, Paschalis C. Sofotasios, \emph{Senior Member, IEEE}\thanks{P. C. Sofotasios is with the Department of Electrical and Computer Engineering, Khalifa
University of Science and Technology, 127788 Abu Dhabi, UAE, and also with the Department of Electronics and Communications Engineering, Tampere University of Technology, 33101 Tampere, Finland, (e-mail:
p.sofotasios@ieee.org)}, Sami Muhaidat, \emph{Senior Member, IEEE}\thanks{S. Muhaidat is with the Department of Electrical and Computer Engineering,
Khalifa University of Science and Technology, Abu Dhabi 127788,
UAE, and also with the Institute for Communication Systems, University of
Surrey, Guildford GU2 7XH, U.K. (e-mail: muhaidat@ieee.org).}, and Simon L. Cotton, \emph{Senior Member, IEEE}\thanks{S. L. Cotton is with Centre for Wireless Innovation,
ECIT Institute, Queen's University Belfast, Belfast BT3 9DT, U.K. (e-mail: simon.cotton@qub.ac.uk).}}\protect}
\maketitle
\begin{abstract}
Capitalizing on the recently proposed Fisher-Snedecor $\mathcal{F}$ composite fading model, in this letter, we investigate the sum of independent but not identically distributed (i.n.i.d.) Fisher-Snedecor $\mathcal{F}$ variates. First, a novel closed-form expression is derived for the moment generating function of the instantaneous signal-to-noise ratio. Based on this, the corresponding probability density function and cumulative distribution function of the sum of i.n.i.d. Fisher-Snedecor $\mathcal{F}$ variates are derived, which are subsequently employed in the analysis of multiple branch maximal-ratio combining (MRC). Specifically, we investigate the impact of multipath and shadowed fading on the outage probability and outage capacity of MRC based receivers. In addition, we derive exact closed-form expressions for the average bit error rate of coherent binary modulation schemes followed by an asymptotic analysis which provides further insights into the effect of the system parameters on the overall performance. Importantly, it is shown that the effect of multipath fading on the system performance is more pronounced than that of shadowing.
\end{abstract}
\IEEEpeerreviewmaketitle
\section{Introduction}
\IEEEPARstart{T}{he} sum of random variables (RVs) finds numerous important applications in wireless communication systems, such as in diversity combining. In particular, maximal-ratio combining (MRC) is considered one of the most efficient diversity techniques that takes advantage of fading to improve system performance \cite{sim}. In this context, a statistical characterization of independent but not identically distributed (i.n.i.d.) $\kappa$-$\mu$ shadowed RVs was addressed in \cite{Paris}. The authors in \cite{OSBadarneh} investigated the performance of $L$-branch MRC receivers under generalized $\eta$-$\mu$ fading with imperfect channel estimation. In \cite{Bithas2007}, the authors analyzed the performance of MRC, equal gain combining (EGC), selection combining (SC) and switch and stay combining (SSC) diversity receivers operating over the composite $K_G$ fading channels. A novel analytical expression for the probability density function (PDF) of the sum of two correlated, dissimilar Nakagami-$0.5$ RVs was derived in \cite{Beaulieu2017}, while the author in \cite{DLee} derived the exact and approximate PDF and cumulative distribution function (CDF) of dual selection with MRC over nonidentical imperfect channel estimation.

It is recalled that composite fading models outperform conventional fading models due to their ability to characterize the simultaneous occurrence of multipath fading and shadowing. Based on this, the authors in \cite{Fisher} proposed the Fisher-Snedecor $\mathcal{F}$ composite fading model which was shown to provide accurate modeling of channel measurements obtained in the context of wearable communications. Based on the empirical data presented therein, it was also shown that this model provided better fit compared to the commonly used $K_G$ fading model, in addition to the added benefit of its algebraic representation which is more tractable. Motivated by this, in the present work, the sum of i.n.i.d. Fisher-Snedecor $\mathcal{F}$ variates is investigated and subsequently employed in the analysis of diversity receivers. To this end, novel analytical expressions for the PDF, CDF, and moment generating function (MGF) are derived in closed-form. These expressions are then used to evaluate the performance of an MRC receiver in terms of outage probability (OP), outage capacity (OC), and average bit error rate (BER) of coherent binary modulation schemes. In addition, a corresponding asymptotic analysis is carried out from which the system diversity gain is determined along with additional insights into the overall system performance.
\section{MGF of Fisher-Snedecor $\mathcal{F}$ Variates}\label{sec:1}
The PDF of the instantaneous signal-to-noise ratio (SNR), $\gamma_{\ell}$, at the $\ell$-th branch of MRC receiver operating under a Fisher-Snedecor $\mathcal{F}$ composite fading channel can be expressed as \cite{Fisher}
\begin{align}\label{pdfr}
f_{\gamma_\ell}(\gamma)={{m_{\ell}^{m_{\ell}}}(m_{s_{\ell}}{\bar{\gamma}_{\ell}})^{m_{s_{\ell}}}\over B(m_{\ell},m_{s_{\ell}})}{\gamma^{m_{\ell}-1}\over\left(m_{\ell} \gamma+m_{s_{\ell}}{\bar{\gamma}_{\ell}}\right)^{m_{\ell}+m_{s_{\ell}}}},
\end{align}
where $m_{\ell}$ and $m_{s_{\ell}}$ denote the fading severity and shadowing parameters, respectively, ${\bar{\gamma}_{\ell}}=\mathbb{E}[\gamma_{\ell}]$ is the mean SNR with $\mathbb{E}[\cdot]$ denoting expectation, and $B(\cdot,\cdot)$ is the beta function \cite[Eq. (8.384.1)]{i:ryz}. The flexibility of the Fisher-Snedecor $\mathcal{F}$ fading model is evident by the fact that it comprises as special cases the Nakagami-$m$ distribution ($m_{s_{\ell}}\rightarrow\infty$, $m_{\ell}=m)$, Rayleigh distribution ($m_{s_{\ell}}\rightarrow\infty$, $m_{\ell}=1$), and one-sided Gaussian distribution ($m_{s_{\ell}}\rightarrow\infty$, $m_{\ell}=0.5$).

The MGF of the Fisher-Snedecor $\mathcal{F}$ distribution was addressed in \cite[Eq. (10)]{Fisher}. However, the algebraic form of the proposed expression renders it inconvenient for the analysis of several scenarios of interest. In what follows, we derive an alternative, closed-form analytical expression for the MGF, which facilitates the derivation of the PDF and CDF of the sum of Fisher-Snedecor $\mathcal{F}$ variates. To this end, by recalling that the MGF is defined as $\mathcal{M}(t)\triangleq\int_{0}^{\infty}\exp{(-x t)}f_{X}(x)dx$, we first represent the exponential function in terms of Meijer's G-function \cite[Eq. (8.4.3.1)]{a:pru} and use the PDF given in (\ref{pdfr}). Then, with the aid of \cite[Eq. (7.811.5)]{i:ryz} and \cite[Eq. (9.31.2)]{i:ryz}, the following analytical expression is deduced
\begin{align}\label{mgf2}
\mathcal{M}_{\gamma_{\ell}}(t)&={1\over \Gamma(m_{\ell})\Gamma(m_{s_{\ell}})}\mathrm{G}_{2,1}^{1,2}\left[{m_{\ell}\over m_{s_{\ell}}\bar{\gamma}_{\ell}t}\left\vert \begin{matrix} 1-m_{s_{\ell}},1\\ m_{\ell}\end{matrix}\right.\right].
\end{align}
Also, the MGF in (\ref{mgf2}) can be written in terms of Tricomi's confluent hypergeometric function (also called confluent hypergeometric function of the second kind) using \cite[Eqs. (8.2.2.14)/(8.4.46.1)]{a:pru}. It is noted here that the MGF of the Nakagami-$m$ distribution can be deduced from (\ref{mgf2}) when $m_{s_{\ell}}\rightarrow\infty$ and $m_{\ell}=m$. As such, using \cite[Eq. (8.2.2.12)]{a:pru}, (\ref{mgf2}) reduces to
 \begin{align}\label{mgf3}
\mathcal{M}_{\gamma}(t)={1\over \Gamma(m)}\mathrm{G}_{1,1}^{1,1}\left[{\bar{\gamma}\over m}t\left\vert \begin{matrix} 1-m\\ 0\end{matrix}\right.\right],
\end{align}
which upon use of \cite[Eq. (8.4.2.5)]{a:pru}, reduces to\cite[Eq. (2.22)]{sim}.

\section{Sum of Fisher-Snedecor $\mathcal{F}$ Variates}\label{sec:2}
\begin{proposition}
Let us consider $\gamma_{\ell}\sim\mathcal{F}\left(\bar{\gamma_{\ell}},m_{\ell},m_{s_{\ell}}\right)$, $\ell=1,\ldots,L$, where all RVs follow i.n.i.d. Fisher-Snedecor $\mathcal{F}$ distributions. The PDF of the sum $\gamma = \sum_{\ell = 1}^{L}  \gamma _{\ell}$ is obtained as
\begin{align}\label{pdfmrcfin}
 {f_\gamma }(\gamma ) = {\gamma^{\sum\limits_{\ell=1}^{L}m_{\ell}-1}\over\Gamma\left(\sum\limits_{\ell=1}^{L}m_{\ell}\right)}\left[ {\prod\limits_{\ell = 1}^{L}}\left({m_{\ell}\over m_{s_{\ell}}\bar{\gamma}_{\ell}}\right)^{m_{\ell}}  {\Gamma(m_{\ell}+m_{s_{\ell}})\over\Gamma ({m_{s_\ell}})} \right]\cr\times F _B^{(L)} \Bigl( {m_1+m_{s_{1}}},{m_2+m_{s_{2}}}, \ldots ,{m_{L}+m_{s_{L}}},{m_1},{m_2},\cr \ldots{m_{L}}; {\sum\limits_{\ell = 1}^{L} {{m_\ell}} ;{{ - {m_1}} \over {m_{s_{1}}{{\bar \gamma }_1}}}\gamma,{{ - {m_2}} \over {m_{s_{2}}{{\bar \gamma }_2}}}\gamma , \ldots ,{{ - {m_{L}}} \over {m_{s_{L}}{{\bar \gamma }_{L}}}}\gamma } \Bigr),\,\,
 \gamma \ge 0,
\end{align}
where $F _B^{(n)}\left(\cdot\right)$ denotes the Lauricella multivariate hypergeometric function \cite[Eq. (1.4.2)]{Srivastava}.

The corresponding CDF can be obtained by performing a term-by-term integration of (\ref{pdfmrcfin}) as
\begin{align}\label{cdfmrcfin}
  {F_\gamma }(\gamma ) = {\gamma^{\sum\limits_{\ell=1}^{L}m_{\ell}}\over\Gamma\left(1+\sum\limits_{\ell=1}^{L}m_{\ell}\right)}\left[ {\prod\limits_{\ell = 1}^{L}}\left({m_{\ell}\over m_{s_{\ell}}\bar{\gamma}_{\ell}}\right)^{m_{\ell}} {\Gamma(m_{\ell}+m_{s_{\ell}})\over\Gamma ({m_{s_\ell}})} \right]\cr  \times F _B^{(L)}\Bigl( {{m_1+m_{s_{1}}},{m_2+m_{s_{2}}}, \ldots ,{m_{L}+m_{s_{L}}},{m_1},{m_2},} \nonumber \\ \ldots{m_{L}}; {1+\sum\limits_{\ell = 1}^{L} {{m_\ell}} ;{{ - {m_1}} \over {m_{s_{1}}{{\bar \gamma }_1}}}\gamma,{{ - {m_2}} \over {m_{s_{2}}{{\bar \gamma }_2}}}\gamma , \ldots ,{{ - {m_{L}}} \over {m_{s_{L}}{{\bar \gamma }_{L}}}}\gamma } \Bigr),\,\,
  \gamma \ge 0,
\end{align}
\end{proposition}
\begin{IEEEproof}
  The proof is provided in the Appendix.
\end{IEEEproof}

Note that (\ref{pdfmrcfin}) and (\ref{cdfmrcfin}) converge if $|{{  {m_1}} \over {m_{s_{1}}{{\bar \gamma }_1}}}\gamma|<1, |{{ {m_2}} \over {m_{s_{2}}{{\bar \gamma }_2}}}\gamma|<1, \ldots ,|{{  {m_{L}}} \over {m_{s_{L}}{{\bar \gamma }_{L}}}}\gamma|<1$. This restriction can be overcome by applying the following transformation to the Lauricella multivariate hypergeometric function \cite[Eq. (7.2.4.36)]{a:pru}
\begin{align}\label{fbident}
&F_B^{(n)}\left(a_{1},\ldots,a_{n},b_{1},\ldots,b_{n};c;x_{1},\ldots,x_{n}\right)=\left[ {\prod\limits_{i = 1}^n {{{(1 - {x_i})}^{ - {b_i}}}} } \right]\cr& \, \times
 F_B^{(n)}\Bigl(c-a_{1},\ldots,c-a_{n},b_{1},\ldots,b_{n};c;{x_{1}\over x_{1}-1},\ldots,{x_{n}\over x_{n}-1}\Bigr).&
\end{align}

It is worth mentioning here that using the relations $\lim\limits_{a\to\infty}{a^{-b}\Gamma(a+b)/\Gamma(a)}=1$ and $\lim\limits_{\min\{|a_{1}|,\ldots,|a_{n}|\}\to\infty}F_B^{(n)}(a_{1},\ldots,a_{n},b_{1},\ldots,b_{n};c;{x_{1}\over a_{1}},\ldots,{x_{n}\over a_{n}})=\Phi_2^{(n)}(b_{1},\ldots,b_{n};c;x_{1},\ldots,x_{n})$ \cite{Srivastava}, (\ref{pdfmrcfin}) and (\ref{cdfmrcfin}) reduce to the Nakagami-$m$ one \cite{VAAalo} when $m_{s_\ell}\rightarrow\infty$.

For the case of i.i.d. Fisher-Snedecor $\mathcal{F}$ variates, the PDF in (\ref{pdfmrcfin}) reduce to \cite[Eq. (7), after correcting some typos]{8359199}
\begin{align}\label{iidpdf}
 & f_{\gamma}(\gamma)= {1\over B\left(Lm ,Lm_{s}\right)}\left({m\over L m_{s}\bar{\gamma}}\right)^{m L}  
  \gamma^{m L-1} \cr&\qquad\qquad\qquad\times\,{}_2F _1\Bigl( L(m+m_{s}),mL;mL,-{  m \over L m_{s}\bar{\gamma}}\gamma \Bigr),&
\end{align}
which can be rewritten, with the help of \cite[Eq. (9.131.1)]{i:ryz} and \cite[9.155.4]{i:ryz}, as follows
\begin{align}\label{iidpdfe}
 & f_{\gamma}(\gamma)= {1\over B\left(Lm ,Lm_{s}\right)}\left({m\over L m_{s}\bar{\gamma}}\right)^{m L}
  \gamma^{m L-1} \cr&\qquad\qquad\qquad\qquad\qquad\times\left( 1+{  m \over L m_{s}\bar{\gamma}}\gamma \right)^{-L(m+m_s)}.&
\end{align}
While the CDF is given by \cite[Eq. (8), after correcting some typos]{8359199}
\begin{align}\label{iicpdf}
 & F_{\gamma}(\gamma)= {\Gamma(Lm+Lm_{s})\over\Gamma (Lm_{s})\Gamma\left(1+mL\right)}\left({m\over L m_{s}\bar{\gamma}}\right)^{mL}  
  \gamma^{mL} \cr&\qquad\qquad\times\,{}_2F _1\Bigl(L( m+m_{s}),mL;1+mL,-{  m \over L m_{s}\bar{\gamma}}\gamma \Bigr),&
\end{align}
whereby ${}_2F _1(\cdot)$ represents the Gauss hypergeometric function \cite[Eq. (9.100)]{i:ryz}. It is noted here that when $|{{  {m}} \over {m_{s}{{\bar \gamma }}}}\gamma|>1$, the transformation in (\ref{fbident}) can be used. Also, when $m_{s}\rightarrow\infty$, using \emph{(i)} $\lim\limits_{a\to\infty}{a^{-b}\Gamma(a+b)/\Gamma(a)}=1$, and \emph{(ii)} $\lim\limits_{|a|\to\infty}{}_2F _1\left(a,b;c;{z\over a}\right)={}_1F _1\left(b,c;z\right)$ \cite{Srivastava}, (\ref{iidpdf}) reduces to the Nakagami-$m$ one \cite{sim} after using \cite[Eq. (9.210.1)]{i:ryz} and \cite[Eq. (9.215.1)]{i:ryz}. In addition, using \emph{(i)} and \emph{(ii)}, (\ref{iicpdf}) reduces to the Nakagami-$m$ one after using \cite[Eq. (9.210.1)]{i:ryz}, \cite[Eq. (8.351.2)]{i:ryz}, and \cite[Eq. (8.356.3)]{i:ryz}.

To the best of the authors' knowledge, equations (\ref{pdfmrcfin}), (\ref{cdfmrcfin}), (\ref{iidpdf}) and (\ref{iicpdf}) have not been previously reported to the open technical literature.
\section{Performance of MRC Receiver}\label{sec:3}
In this section, we analyze the performance of an $L$-branch MRC receiver operating under Fisher-Snedecor $\mathcal{F}$ composite fading in terms of the OP, OC and BER for coherent binary modulations.
\subsection{Outage Probability}
It is recalled that $P_{out}\triangleq \Pr[0\leq\gamma<\gamma_{{\rm th}}]=\int_{0}^{\gamma_{{\rm th}}} f_{\gamma}(\gamma)d\gamma$; therefore, the corresponding OP is readily deduced as
\begin{align}
P_{out}= F_{\gamma}(\gamma_{{\rm th}}),
\end{align}
where $F_{\gamma}(\gamma)$ is given in (\ref{cdfmrcfin}).

\subsubsection{Diversity Gain} The diversity gain or diversity order refers to the increase in the slope of the OP, $P_{out}$, versus the average SNR. Thus, at high average SNR value, the OP, $P_{out}$, may be closely approximated by $P_{out} \simeq {\bar{\gamma}}^{-G_d}$, whereby the exponent $G_d$ represents the diversity gain.

The asymptotic OP can be obtained when $\bar{\gamma}_{\ell}\rightarrow\infty$ for all $\ell$. As such, making use of the identity $F_B^{(n)}(a_{1},\ldots,a_{n},b_{1},\ldots,b_{n};c;\underbrace{0,\ldots,0}_{n})=1$, an asymptotic expression for the OP can be represented as follows:
\begin{align}\label{opasymp}
P_{out}\simeq {\gamma_{{\rm th}}^{\sum\limits_{\ell=1}^{L}m_{\ell}}\over\Gamma\left(1+\sum\limits_{\ell=1}^{L}m_{\ell}\right)}\left[ {\prod\limits_{\ell = 1}^{L}}\left({m_{\ell}\over m_{s_{\ell}}\bar{\gamma}_{\ell}}\right)^{m_{\ell}} {\Gamma(m_{\ell}+m_{s_{\ell}})\over\Gamma ({m_{s_\ell}})} \right].
\end{align}
It is evident from (\ref{opasymp}) that the diversity gain $G_d$ is proportional to ${L}$ and $m$. It is also noted that for the i.i.d. case, $G_d=m L$.
\subsection{Outage Capacity}
The OC is an important statistical measure used to quantify the spectral efficiency over fading channels. It can defined as the probability that the instantaneous capacity $C_{\gamma}$ falls below a certain specified threshold $C_{{\rm th}}$, that is, $C_{out}\triangleq \Pr[0\leq C_{\gamma}<C_{{\rm th}}]$, where $C_{\gamma}=W\log_{2}(1+\gamma)$ and $W$ denotes the signal's transmission bandwidth over AWGN. Thus, the OC can be written in terms of the CDF as follows
\begin{align}
C_{out}= F_{\gamma}(2^{C_{{\rm th}}\over W}-1).
\end{align}
To this effect, the OC in MRC receivers under $\mathcal{F}$ composite fading conditions is readily obtained using the derived expression in (\ref{cdfmrcfin}).
\subsection{Average Bit Error Rate}
In flat fading environments, the average BER of most coherent modulation techniques can be calculated as \cite{sim}
\begin{align}\label{berb}
{\overline P_b} = \int\limits_0^\infty Q\left(\sqrt{2\lambda\gamma}\right){f_\gamma }(\gamma )d\gamma={1\over\pi}\int\limits_{0}^{\pi/2}\mathcal{M}_{\gamma}\left(\lambda\over\sin^{2}\phi\right)d\phi,
\end{align}
where $Q(\cdot)$ is the well-known Gaussian Q-function and $\lambda$ is a dependent-modulation constant. For binary phase shift keying (BPSK) $\lambda=1$, while for binary frequency shift keying (BFSK) $\lambda=0.5$, and for BFSK with minimum correlation $\lambda=0.715$. Based on this and utilizing (\ref{mgf2}) and (\ref{berb}), the average BER ${\overline P_b}$ for the considered receiver under $\mathcal{F}$ fading can be expressed as
\begin{align} \label{berbin}
{\overline P_b} = {1\over\pi}\prod\limits_{\ell = 1}^{L}\int\limits_{0}^{\pi/2}{\mathrm{G}_{2,1}^{1,2}\left[{m_{\ell}\over \lambda m_{s_{\ell}}\bar{\gamma}_{\ell}}\sin^{2}\phi\left\vert \begin{matrix} 1-m_{s_{\ell}},1\\ m_{\ell}\end{matrix}\right.\right]\over \Gamma(m_{\ell})\Gamma(m_{s_{\ell}})}d\phi.
\end{align}

By performing the change of variable $y=\sin^{2}\phi$ and using \cite[2.24.2.2]{a:pru}, the average BER ${\overline P_b}$ can be obtained in closed-form as
\begin{align}\label{aberbsp}
&{\overline P_b}={1\over2\sqrt{\pi}}\prod\limits_{\ell = 1}^{L}{\mathrm{G}_{3,2}^{1,3}\left[{m_{\ell}\over\lambda m_{s_{\ell}}\bar{\gamma}_{\ell}}\left\vert \begin{matrix} {1\over2},1-m_{s_{\ell}},1\\ m_{\ell},0\end{matrix}\right.\right]\over \Gamma(m_{\ell})\Gamma(m_{s_{\ell}})},&
\end{align}
which has a tractable algebraic representation. To this effect, the asymptotic average BER is obtained by assuming $\bar{\gamma}_{\ell}\rightarrow\infty$ for all $\ell$. Using (\ref{aberbsp}) and utilizing \cite[Eq. (8.3.2.21)]{a:pru}, and \cite[Eq. (1.8.5)]{kilbas}, yields
\begin{align} \label{berasym}
&{\overline P_b}\simeq{1\over2\sqrt{\pi}}{\prod\limits_{\ell = 1}^{L}}\left({m_{\ell}\over\lambda m_{s_{\ell}}}\right)^{m_{\ell}} \left({1\over\bar{\gamma}_{\ell}}\right)^{m_{\ell}} { \Gamma(m_{\ell}+m_{s_{\ell}})\Gamma({1\over2}+m_{\ell})\over\Gamma(m_{s_{\ell}})\Gamma(1+m_{\ell})}.&
\end{align}
It is evident that the diversity gain $G_d$ (for the i.n.i.d. case) is proportional to ${L}$ and $m$, whereas for the i.i.d. case $G_d=m L$. Furthermore, the derived asymptotic expressions in (\ref{opasymp}) and (\ref{berasym}) have a convenient algebraic representation that renders them convenient to handle both analytically and numerically.

\section{Numerical Results and Discussions}\label{sec:4}
This section presents numerical and simulation results for the considered MRC receiver with $L$ branches. It can immediately be observed that a tight agreement between numerical and simulation results exists, which verifies the validity of the derived expressions. In addition, the asymptotic approximations match well with both the numerical and simulation results at high SNR values. The results for the Rayleigh case are also included as a benchmark.

\begin{figure}[t!]
\centering
 \includegraphics[width=0.8\columnwidth]{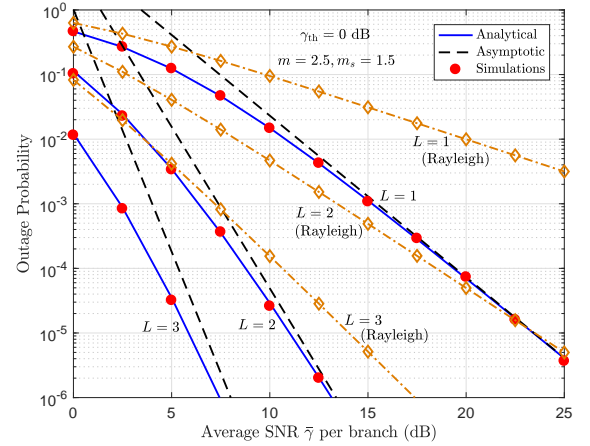}
\caption{Outage probability versus average SNR $\bar{\gamma}$ per branch.}%
\label{fig:op}
\end{figure}

In Fig. \ref{fig:op}, the OP performance is plotted as a function of the average SNR $\bar{\gamma}$ per branch for i.i.d. Fisher-Snedecor $\mathcal{F}$ variates with $m=2.5$ (moderate), $m_s=1.5$ (heavy), and $\gamma_{{\rm th}}=0~{\rm dB}$. The results show that the difference between $L = 1$ and $L = 2$ is significant as about half of the power is required in the latter to achieve a target BER across all SNR regimes.

\begin{figure}[t]
\centering
\includegraphics[width=0.8\columnwidth]{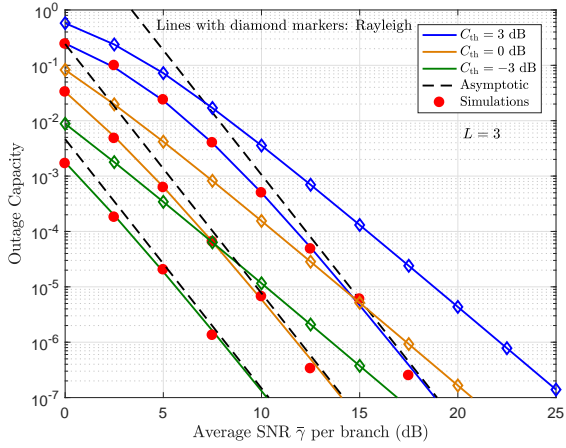}
\caption{Outage capacity versus average SNR $\bar{\gamma}$ per branch.}
\label{fig:oc}
\end{figure}

The OC is plotted in Fig. \ref{fig:oc} as a function of the average SNR $\bar{\gamma}$ per branch for different values of $C_{{\rm th}}$ with $L=3$, $m=1.5$, and $m_s=1.25$. It is noticed that the SNR gains for extra branches are similar from $L = 1$ to $L =2$ and from $L = 2$ to $L = 3$. Also, it can be concluded that three branches are sufficient to achieve low OC even at low SNR regimes, which is feasible as it does not come at a cost of dramatic complexity increase.

\begin{figure}[t]
\includegraphics[width=0.8\columnwidth]{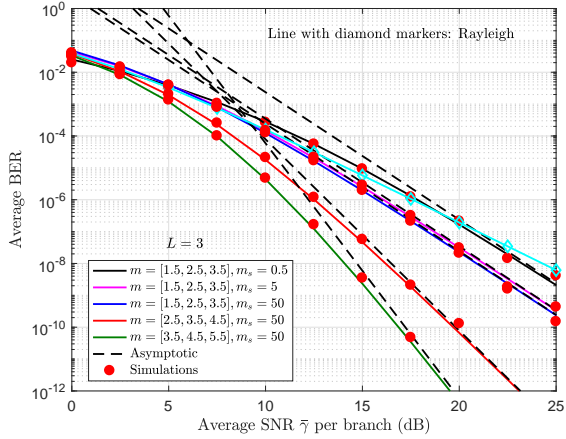}
\centering\caption{Average BER for BPSK versus average SNR $\bar{\gamma}$ per branch.}%
\label{fig:ber}
\end{figure}

In Fig. \ref{fig:ber}, the performance of the average BER for coherent BPSK modulation and i.n.i.d. triple-branch MRC is depicted. The results show the influence of shadowing $m_s=0.5$ (heavy), $m_s=5$ (moderate), and $m_s=50$ (light) on the average BER. It is apparent that the average BER slightly improves as $m_s$ increases. In addition, the results show that the impact of the fading parameter $m$ on the system performance is more pronounced than that of $m_s$. This is because the diversity order of the system is proportional to $m$, as shown in (\ref{berasym}).
\section{Conclusions}
A new closed-form expression for the MGF of Fisher-Snedecor $\mathcal{F}$ distribution has been obtained. Capitalizing on this, novel closed-form expressions for the PDF and CDF of the sum of i.n.i.d. Fisher-Snedecor $\mathcal{F}$ variates were derived. Using the new expressions, useful theoretical and technical insights into the performance of an $L$-branch MRC receiver operating over Fischer-Scnedecor $\mathcal{F}$ fading have been presented in terms of OP, OC, and average BER.
\appendix[Proof of Proposition 1]
The PDF of $\gamma$ can be found using the inverse Laplace transform, namely
 \begin{align}\label{invlap}
   f_{\gamma}(\gamma)={1\over2\pi j}\int\limits_{\mathbb{C}}\mathcal{M}{_\gamma }(t){e^{\gamma t}}dt,
 \end{align}
where $\mathcal{M}{_\gamma }(t)$ is the MGF of $\gamma$, which can be written as
\begin{align}\label{mgftot}
\mathcal{M}{_\gamma }(t) = \prod\limits_{\ell = 1}^{L} \mathcal{M}_{\gamma _{\ell}}(t).
\end{align}
To obtain the desired results, we rewrite $\mathcal{M}_{\gamma _{\ell}}(t)$ in (\ref{mgf2}) using \cite[Eq. (9.31.5)]{i:ryz} as
\begin{align}\label{mgfdes}
\mathcal{M}_{\gamma_{\ell}}(t)&={\left({m_{\ell}\over m_{s_{\ell}}\bar{\gamma}_{\ell}t}\right)^{m_{\ell}}\over \Gamma(m_{\ell})\Gamma(m_{s_{\ell}})}\mathrm{G}_{2,1}^{1,2}\left[{m_{\ell}\over m_{s_{\ell}}\bar{\gamma}_{\ell}t}\left\vert \begin{matrix} 1-m_{\ell}-m_{s_{\ell}},1-m_{\ell}\\0\end{matrix}\right.\right].
\end{align}
In the case of i.i.d. Fisher-Snedecor $\mathcal{F}$ RVs, the MGF of $\gamma$ can be obtained using (\ref{iidpdf}) and \cite[Eq. (3.37.1.8)]{pru4}.
With the help of (\ref{mgfdes}) and the definition of the Meijer's G-function \cite[Eq. (9.301)]{i:ryz}, (\ref{mgftot}) can be rewritten as
\begin{align*}
& {\mathcal{M}_\gamma }(t) = \left[ {\prod\limits_{\ell = 1}^{L} {{1 \over {\Gamma ({m_\ell})\Gamma ({m_{s_\ell}})}}} } \left({m_{\ell}\over m_{s_{\ell}}\bar{\gamma}_{\ell}t}\right)^{m_{\ell}} \right]{\left( {{1 \over {2\pi j}}} \right)^{L}}\int\limits_{\mathbb{C}_{1}} {\int\limits_{\mathbb{C}_{2}} \cdots } \cr & \,\,\,\,\,\,\,\,\,\,\,\,\,\,\,\,\,\,\int\limits_{\mathbb{C}_{L}} {\left\{ {\prod\limits_{\ell = 1}^{L} \Gamma(-s_{\ell})\Gamma(m_{\ell}+s_{\ell})\Gamma(m_{\ell}+m_{s_{\ell}}+s_{\ell})
} \right\}} \cr & \,\,\,\,\,\,\,\,\,\,\,\,\,\,\,\,\,\,\,\,\,\,\,\,\,\,\,\,\,\,\times \left({m_{\ell}\over m_{s_{\ell}}\bar{\gamma}_{\ell}t}\right)^{s_{\ell}}d{s_1}d{s_2} \ldots d{s_{L}}.\qquad\qquad\qquad\hbox{(19)}&
\end{align*}
Next, substituting (19) into (\ref{invlap}), the PDF of $\gamma$ can be rewritten as follows:
\begin{align*}
 & {f_\gamma }(\gamma ) =  \left[ {\prod\limits_{\ell = 1}^{L} {{1 \over {\Gamma ({m_\ell})\Gamma ({m_{s_\ell}})}}} } \left({m_{\ell}\over m_{s_{\ell}}\bar{\gamma}_{\ell}}\right)^{m_{\ell}} \right]{\left( {{1 \over {2\pi j}}} \right)^{L}}\int\limits_{\mathbb{C}_{1}} {\int\limits_{\mathbb{C}_{2}} \cdots } \cr & \,\,\,\,\,\,\,\,\,\,\,\,\,\,\,\,\,\,\int\limits_{\mathbb{C}_{L}} {\left\{ {\prod\limits_{\ell = 1}^{L} \Gamma(-s_{\ell})\Gamma(m_{\ell}+s_{\ell})\Gamma(m_{\ell}+m_{s_{\ell}}+s_{\ell})
} \right\}} \cr&\times \left({m_{\ell}\over m_{s_{\ell}}\bar{\gamma}_{\ell}}\right)^{s_{\ell}}\underbrace{ \left( {{1 \over {2\pi j}}\int\limits_{\mathbb{C}} {{t^{ - \sum\limits_{\ell = 1}^{L} {({m_\ell} + {s_\ell})} }}{e^{\gamma t}}dt} }  \right)}_{\mathcal{I}_{1}}d{s_1}d{s_2} \ldots d{s_{L}}.\hbox{(20)}&
\end{align*}
After making the change of variable $y=-\gamma t$ in (20) and solving the integral $\mathcal{I}_{1}$ using \cite[Eq. (8.315.1)]{i:ryz}, it follows that
\begin{align*}
  {f_\gamma }(\gamma ) =  \left[ {\prod\limits_{\ell = 1}^{L}} \left({m_{\ell}\over m_{s_{\ell}}\bar{\gamma}_{\ell}}\right)^{m_{\ell}} \right]\gamma^{\left(\sum\limits_{\ell=1}^{L}m_{\ell}\right)-1}
 {\left( {{1 \over {2\pi j}}} \right)^{L}}\int\limits_{\mathbb{C}_{1}} {\int\limits_{\mathbb{C}_{2}} \cdots }\cr \int\limits_{\mathbb{C}_{L}} {1\over\Gamma\left(\sum\limits_{\ell=1}^{L}(m_{\ell}+s_{\ell})\right) } \left\{ \prod\limits_{\ell = 1}^{L} {\Gamma(-s_{\ell})\Gamma(m_{\ell}+s_{\ell})\over\Gamma ({m_\ell})\Gamma ({m_{s_\ell}})}
 \right. \nonumber\\
    \left.\times\Gamma(m_{\ell}+m_{s_{\ell}}+s_{\ell})\left({m_{\ell}\gamma\over m_{s_{\ell}}\bar{\gamma}_{\ell}}\right)^{s_{\ell}} \right\}  d{s_1}d{s_2} \ldots d{s_{L}}.\qquad\hbox{(21)}
\end{align*}
Notably, the multiple Barnes-type contour integrals in (21) can be represented in terms of Lauricella multivariate hypergeometric function \cite[Eq. (1.10.10)]{AMMathai}, which completes the proof.
%
%
\bibliographystyle{IEEEtran}
\bibliography{prodff}
\end{document}